\def\be{\begin{equation}}
\def\ee{\end{equation}}
\def\bea{\begin{eqnarray}}
\def\eea{\end{eqnarray}}
\def\Dt{\Delta t}
\def\Dx{\Delta x}
\documentclass[twocolumn,twoside,preprintnumbers,amsmath,amssymb,showkeys,showpacs,floatfix]{revtex4}
\usepackage{epsfig,graphicx,fancyhdr}

\begin{document}
\title{Anomalous diffusion of pions at RHIC}

\author{M. Csan\'ad$^1$, T. Cs\"org\H{o}$^{2,3}$ and M. Nagy$^1$ }

\affiliation{ $^1$ Dept. Atomic Phys., ELTE,
     H - 1117 Budapest, P\'azm\'any P. s. 1/A, Hungary
    $^2$Instituto de F\'\i sica Te\'orica - UNESP,
    Rua Pamplona 145, 01405-900 S\~ao Paulo, SP, Brazil, \\
         $^3$ MTA KFKI RMKI, H - 1525 Budapest 114, P.O.Box 49, Hungary \\
    }

\begin{abstract}
After pointing out the difference between normal and anomalous diffusion,
we consider a hadron resonance cascade (HRC) model simulation for
particle emission at RHIC and point out that rescattering in
an expanding hadron resonance gas leads
to a heavy tail in the source distribution.
The results are compared to recent PHENIX measurements of the tail of the
particle emitting source in Au+Au collisions at RHIC.
In this context, we show how can one distinguish experimentally the
 anomalous diffusion of hadrons
from a second order QCD phase transition.
\end{abstract}

\vskip -1.35cm

\maketitle

\thispagestyle{fancy}

\setcounter{page}{1}

\bigskip
\begin{verse}
{\it `` It does  not  make  any difference  \\
how  beautiful your guess is. \\
It does  not  make  any difference \\
 how  smart you are, \\
who made the guess,  or  what  his name is - \\
if it disagrees with experiment,
it is wrong."           \\
\hspace{2cm} \\
\hspace{4cm}    /R.P. Feynman/}
\end{verse}
\section{INTRODUCTION}
\label{s:intro}
Various new techniques are being developed in a vibrant and inspiring, sometimes
challenging and puzzling sub-field, named recently by Lednicky~\cite{Lednicky:2005af}
 as {\it correlation femtoscopy}. A series of inspiring and stimulating recent
reviews~\cite{Padula:2004ba,Csorgo:2005gd,Lisa:2005dd,Lisa:2005js,Hwa:2007rp}
re-connected femtoscopy to the search of new phases of QCD as well as other new,
sometimes unexpected, sometimes puzzling expectations and observations.
One of these new experimental results has been a recent PHENIX measurement
of source images in Au+Au collisions
at $\sqrt{s_{NN}} = 200$ GeV that found evidence for a non-Gaussian structure
and a heavy (larger than Gaussian) tail in the source distribution
for pions ~\cite{PHENIX-imaging}.
  This points to an interesting new direction, going beyond investigating only the
means and the variances of the source distributions on the scales of femtometers.

Resonance decays are known to be able to produce long, non-Gaussian tails,
because some of the resonances have large decay times as compared to the
characteristic 4-5 fm/c source sizes extracted from interferometric
measurements in Au+Au collisions at RHIC.
In fact, a smaller and smaller fraction of resonances have larger and larger
life-times, and this effect was considered first by Bia{\l}as~\cite{Bialas:1992ca}
who argued that even a cut power-law correlation functions may appear,
due to the fact that the decay times of pion producing resonances
have a broad probability distribution.
Another possible conventional source for such a heavy tail of the
source of pions might be the elastic rescattering of the produced hadrons:
as the hadron gas expands, the system becomes cooler and more and more diluted,
hence the mean free path becomes larger and larger.
In many hydrodynamical calculations, an idealized freeze-out process
is assumed, when the mean free path suddenly jumps from 0 (the hydrodynamic limit) to infinity (the freeze-out limit). More realistically, the mean free path
diverges to infinity in a finite time interval, and rescattering in a time
dependent mean free path system is known to lead to
new phenomena, which has been studied in great detail under the
name of {\it anomalous diffusion} in other branches of physics.
One of the experimentally observed characteristics of such an anomalous diffusion pattern
was the appearance of approximately power-law shaped tails in the coordinate space
distributions, which is to be contrasted to the Gaussian, strongly decaying
tails observed in normal diffusion or in Brownian motion.
We shall explore the phenomena of anomalous diffusion in the context of heavy ion physics,
first pointing out its general mathematical properties and its relations
to L\'evy source distributions, following the review of Klafter and Metzler
on anomalous diffusion~\cite{MK-rep}.
Then we shall explore anomalous diffusion of pions, kaons and protons
in Au+Au collisions at $\sqrt{s_{NN}} = 200 $ GeV
colliding energies using the simplest possible tool, namely the conventional
Hadronic Resonance Cascade (HRC) model of Tom Humanic~\cite{Humanic:2005ye}.
First we compare the results of this HRC simulation with PHENIX data.
Then we investigate the sensitivity of the characteristics of the simulation
 for various experimentally available controls,
like the selection of the centrality
class of the events or the selection of a transverse momentum region of the pair.
Finally we summarize and conclude.  As a starting point, let us consider a simplified version of the presentation in ref.~\cite{MK-rep},
to review the equations of normal and anomalous diffusions in the same framework,
based on a master equation approach.
\section{Normal Diffusion}
\label{ss:normal-diff}

Normal diffusion corresponds to a physical process when
one investigates the motion of a test particle in  a medium,
which has some grid like structure with fixed (say, $\Dx$) lattice constant,
(equidistant cells) and jumps (of a fixed frequency) are allowed to nearest
neighboring cells. Let us for simplicity consider a one dimensional, normal diffusion
process in the framework of master equation approach.
The time is denoted by $t$. A test particle is
initially located at a given cell, denoted by $j=0$. Different cells
are indexed by the integer $j \in \mathbb{Z}$. The probability distribution
of finding the particle in cell $j$ at time $t$ is denoted by $W_j(t)$.
(Note that $W_j(t)$ can be considered as analogous to the
particle emission function $S(x,t)$,
that we frequently encounter in particle interferometry and femtoscopy.)
Suppose that particles may jump from the given cell to its nearest neighbors,
randomly up and down, at certain regular but small time intervals $\Delta t$.
The continuum limit means $\Delta t \rightarrow 0$, $\Dx\rightarrow 0$ in such a way
that the mean free path (and also the mean collision time) remains constant.

In this situation, the following master equation drives the time evolution in this
material:
\be
W_j(t + \Dt) = \frac{1}{2} W_{j-1}(t) + \frac{1}{2} W_{j+1}(t) .
\ee
We rewrite this discretized form to a continuous form by introducing the
continuous coordinate variable $x$.
We assume that the $\Dx$ step size is infinitesimally small
compared to the overall length-scale of the medium,
and that the time interval between the subsequent jumps
is small as compared to the time duration corresponding to the observation
of the diffusion process. These assumptions yield
the following leading order Taylor expansions:
\be
W_j(t + \Dt)  =  W_j(t) + \Dt \frac{\partial W_j}{\partial t} + {\cal O}(\Dt^2),
\ee
\be
W_{j \pm 1}(t)  =  W(x,t) \pm \Dx \frac{\partial W}{\partial x} +
\frac{(\Dx)^2}{2}\frac{\partial^2 W}{\partial x^2}
+ {\cal O}(\Dx^3),
\ee
and these expressions can be combined to derive the continuum form of the
diffusion equation
\be
\frac{\partial W}{\partial t}  = K_1
\frac{\partial^2 }{\partial x^2} W(x,t) .
\ee
All properties
of this matter are characterized by the diffusion constant
$K_1 = \lim_{\Delta x \rightarrow 0, \Dt \rightarrow 0} \frac{\Dx^2}{2 \Dt}$,
which corresponds to the mean squared displacement per unit time.

The above (normal) diffusion equation can be solved easily by introducing the Fourier-transform
\be
    W(k,t) = \int\, d x \exp(i k x) W(x,t) ,
\ee
that leads to the momentum-space diffusion equation
\be
\frac{\partial W}{\partial t}  = - K_1  k^2 W(k,t) .
\ee
The solution of this equation is a Gaussian function of $k$ that can be converted
back to the coordinate-space distribution to yield the solution of the
normal diffusion equation:
\be
    W(x,t) = \frac{1}{\sqrt{ 4 \pi K_1 t} }
        \exp\left( - \frac{x^2}{ 4 K_1 t}
    \right) ,
    \label{e:diffusion}
\ee
corresponding to normal or Gaussian diffusion of test particles.
The initial condition corresponding to this solution is indeed
\be
    W_0(x) \equiv \lim_{t\rightarrow 0} W(x,t) = \delta(x) \, ,
\ee
corresponding to a localized package inserted in a
material with homogeneous, time independent properties.
Hence in a homogeneous and time independent medium,
the diffusion of point-like initial source yields
a Gaussian source density distribution, and the mean square of this Gaussian,
$R^2 = 2 K_1 t$ increases linearly with increasing time.
This behavior will be contrasted to the more general case, when both the
jump lengths and the jump frequencies have a continuous probability distribution.

\section{Anomalous Diffusion}
\label{ss:anom-diff}

We again follow Metzler and Klafter in deriving anomalous diffusion
from the so-called continuous time random walk models. Suppose that
the jump length and jump time have the probability distribution
$\psi(x,t)$. Then the jump length distribution $\lambda(x)$ and the
waiting time or jump time distribution $w(t)$ reads as
\bea
    \lambda(x) & =  & \int_0^\infty \psi(x,t)\, \mathrm{d}t, \\
    w(t) & =  & \int_{-\infty}^\infty \psi(x,t)\, \mathrm{d}x.
\eea
Suppose that an external force $F(x)$ also influences the motion
of the test particles.
These considerations lead to the generalized or anomalous diffusion
equation, which is a kind of a generalized Fokker-Planck equation
for the phase-space distribution, $W(x,v,t)$:
\be
\frac{\partial W}{\partial t}  + v \frac{\partial W}{\partial x}  +
\frac{F(x)}{m} \frac{\partial W}{\partial v}
= \eta_{\alpha^{\prime}} \null_0 D_t^{1 - \alpha^\prime} L_{FP} W ,
\ee
which contains fractional derivatives and other subtleties
that go well beyond the scope of this presentation.
For the detailed definition and explanation of
this fractional Fokker-Planck equation we refer to
~\cite{MK-rep}. What is important, however, that if the  waiting
time distribution has a Poissonian shape, the exact solution of this
fractional Fokker-Planck equation has a simple form in the momentum space
representation:
\be
    W(k,t) = \exp( - t K^{\alpha} |k|^{\alpha}) .
\ee
This form is the well known characteristic function (Fourier-transform)
of L\'evy stable source distributions~\cite{zol2,zol1,nolan-chap1,nolan-book}.
 Here the parameter  $\alpha$ stands for the
L\'evy index of stability, in general $0 < \alpha \le 2$ for L\'evy
stable source distributions, and parameter $K$ is an anomalous diffusion constant,
$K = \lim_{\Delta x \rightarrow 0, \Dt \rightarrow 0} \frac{\Dx^2}{2 \Dt^{2/\alpha}}$.

Note that L\'evy stable distributions were introduced
to particle interferometry studies recently in
ref.~\cite{csorgo-hegyi-zajc}, based on general, mathematical arguments
like generalized central limit theorems. Anomalous diffusion
is a specific example of a physical process that
under certain conditions detailed in~\cite{MK-rep} satisfies
such generalized central limit theorems: one more rescattering in
the diffusion process does not change the limiting behavior of the source
distribution.

\section{L\'evy stable laws and anomalous diffusion}
\label{ss:anomlevy}

The edification from the previous section is that rescattering in a system
with a time dependent mean free path under certain conditions
(e.g. Poissonian waiting time distributions) leads to a random L\'evy walk
(or anomalous diffusion), instead of Brownian motion (or normal, Gaussian diffusion).
In case of anomalous diffusion or L\'evy walks, the scale parameter grows with time
as  $R^\alpha \propto t$, in contrast to normal diffusion,
that corresponds to the $\alpha = 2 $, $R^2 \propto t$ special case.

\begin{figure}[!htb1]
\begin{center}
\includegraphics*[angle=0, width=7cm]{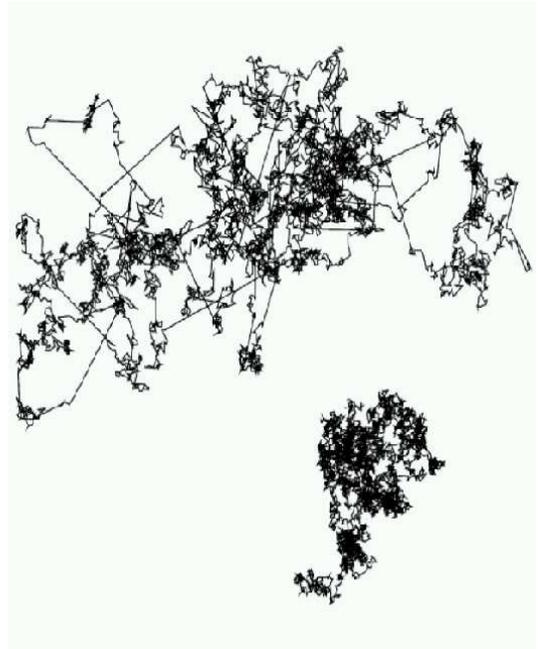}
\end{center}
\caption{
\label{f:anom}
\emph{ \small Simple illustration of the qualitative properties
of normal diffusion (bottom) and anomalous diffusion (top).
In the latter case, large jumps separate various more local parts
of the trajectory. The size of the covered region is larger in case
of the anomalous diffusion than in case of the Gaussian,
normal diffusion.  From ref.~\cite{MK-rep}.}}
\end{figure}

The difference  between normal and anomalous diffusion
is illustrated in Fig.~\ref{f:anom}, reproduced from ref.~\cite{MK-rep}.

Although the mean
displacement or the variance of these distributions might
diverge, one additional step in the anomalous diffusion
does not change the limiting behavior of the process.

It is also interesting to note that the tail of anomalous
diffusions has a power-law structure in coordinate space,
$S(r) \propto r^{-(d+\alpha)}$ for $r \gg R$, where $d$ is
the number of spatial dimensions ($d= 3$ in our world), and
$\alpha$ is  the same L\'evy index of stability as before.
This asymptotics is yet another important property of the L\'evy stable source distributions.

Nature often violates Gaussian universality, mirrored in experimental
results which do not follow Gaussian predictions~\cite{MK-rep}.
The evidence for a non-Gaussian behavior in multivariate
and univariate source distributions has been observed recently
also in high energy heavy ion collisions.
In fact, the first observation of a non-Gaussian correlation
function in Au+Au collisions at RHIC has been made by the STAR
collaboration that utilized an Edgeworth expansion~\cite{STAR-edge}
and quantified the deviation from the Gaussian structure
of the particle emitting source in terms of non-vanishing fourth
order cumulant moments of the squared Fourier-transformed source
distribution. More recently, the PHENIX Collaboration
has applied the imaging method of Brown and Danielewicz~\cite{imaging}
to reconstruct the two-particle relative
coordinate distribution~\cite{PHENIX-imaging}.
In this paper, PHENIX observed a clear deviation from
a Gaussian structure, and pointed to the appearance of
a heavy tail in the two-particle relative coordinate distribution.
In the subsequent part of this manuscript, we investigate
if these PHENIX source function data can be reproduced
with the help of Monte-Carlo models that incorporate the
concept of anomalous diffusion and that are able to
describe the more standard   observable like
single particle spectra and the three dimensional Gaussian fit parameters,
$R_\mathrm{side}$, $R_\mathrm{out}$ and  $R_\mathrm{long}$ of the
measured two-pion correlation functions
in high energy heavy ion collisions.

\section{MONTE-CARLO SIMULATIONS OF ANOMALOUS DIFFUSION OF PIONS}
\label{s:simulations} Heavy ion collisions produce thousands of
particles in a single high energy nucleus-nucleus collision. The
bulk of the particle production, i.e. the momentum distribution
and the correlation patterns of 99 \% of the particles are best
described by hydrodynamical, or hydrodynamically inspired models,
see ref.~\cite{Csorgo:2005gd}
for a recent review.
Now we are interested in the tails of particle production, which might
indicate a deviation from the hydrodynamical behavior. Hence our
attention is turned to Monte-Carlo (MC) simulations.
Two Monte-Carlo models, the Hadronic (or Humanic) Resonance Cascade
Model~\cite{Humanic:2003gs}, (HRC) and the AMPT  model of Zhang, Ko, Li and Lin
~\cite{Zhang:1999bd}
have also demonstrated~\cite{Humanic:2002a,Humanic:2005ye,Ko:2002iz}
their ability to describe single particle spectra,
elliptic flow, and HBT correlation measurements
in Au+Au collisions with $\sqrt{s_{NN}}= 130$ and 200 GeV at RHIC.

\subsection{Selection criteria - comparison with data}
\label{ss:data}
The selection of the MC model was based on the following criteria. We
looked for a conventional hadronic cascade model that
describes single particle spectra, elliptic flow data, and HBT data
(without any puzzles), hence yields a good description of the
hadronic final state, i.e. it yields an acceptable model of S(r).

It also has to be well documented and easy to use, has to work at
CERN SPS as well as at RHIC energies, contain the most important
short and long lived resonances e.g. $\omega$, $\eta$ and $\eta'$
(hence can be used as a realistic model for halo
that appears from the decay products of these  long-lived resonances).
It is important to require that the simulation
takes into account  the rescattering among the hadrons due to
possible development of a power-law tail from the anomalous diffusion,
discussed in the earlier sections.

Finally we utilized the Hadronic Rescattering
Model~\cite{Humanic:2003gs}, as it satisfied all the criteria listed above.
We expect that similar results are obtained with the AMPT model,
but we did not yet investigated the predictions of this code in detail.
The AMPT model is a multi-phase transport model, while the HRC is
a simpler hadronic resonance cascade. When looking for new effects
related to the anomalous diffusion in a time dependent mean free path
environment, we have opted for the simplest possible choice - so that the
results then can be uniquely related to the hadronic final state effects.

\begin{figure}
\begin{center}
\includegraphics[width=1\linewidth]{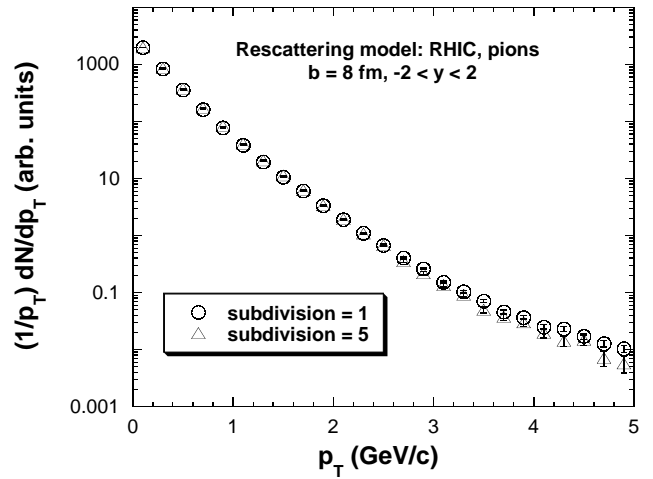}
\caption{\label{f:hrc1} Pion $p_t$ distributions for $l=1$ and
$l=5$, from ref.~\cite{Humanic:2005ye}.
The insensitivity of the
transverse momentum distribution to the number of subdivisions
indicates that the scaling properties of the transport equations
are satisfied by the HRC code. Similar
subdivision tests are satisfied by the HRC code for the elliptic flow
as a function of the transverse momentum $v2(p_t)$, the elliptic flow
as a function of the longitudinal angular variable $\eta$ as $v_2(\eta)$,
see ref.~\cite{Humanic:2005ye} for further details.}
\end{center}
\end{figure}

\begin{figure}
  \includegraphics[width=0.8\linewidth]{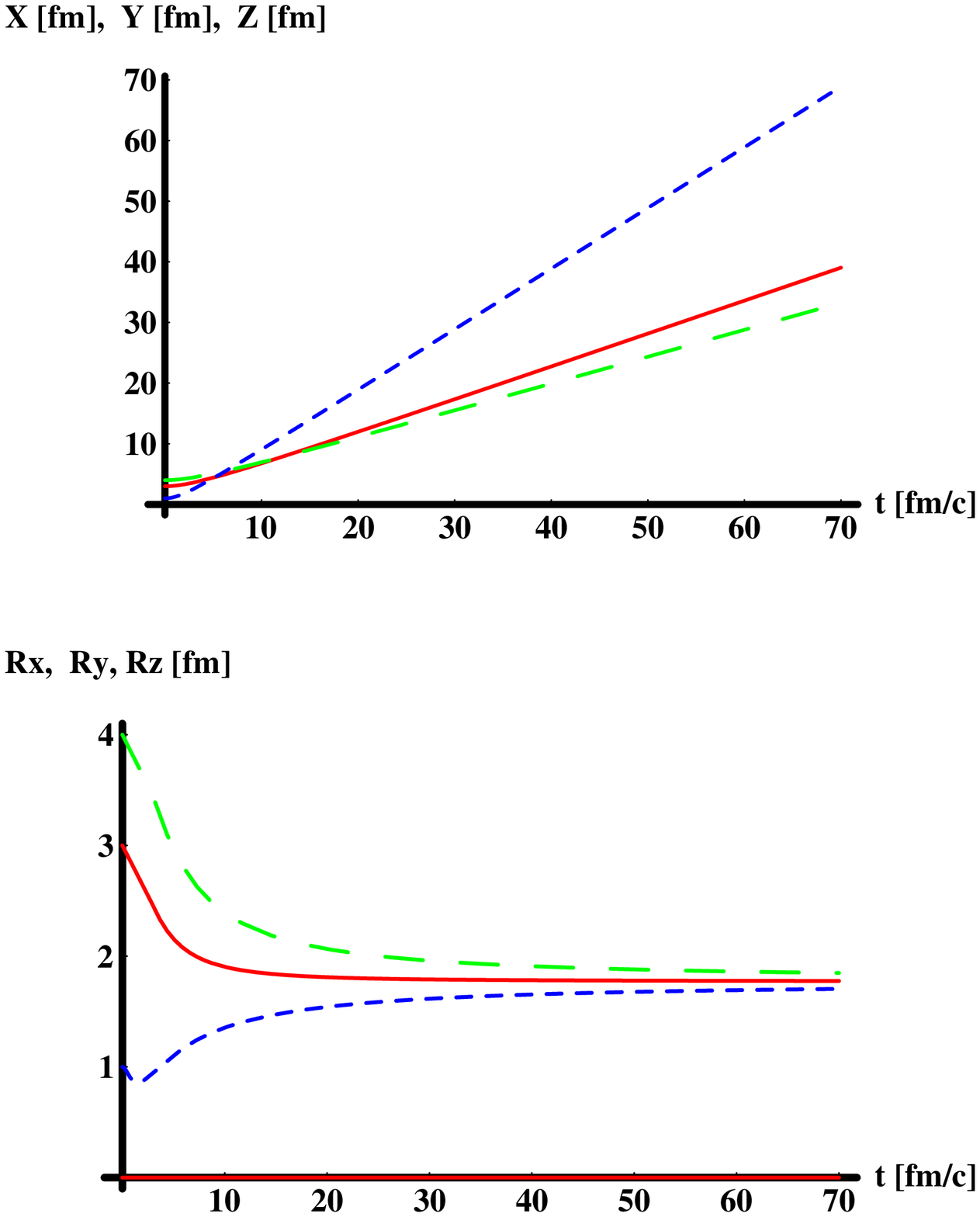}\\
  \includegraphics[width=0.8\linewidth]{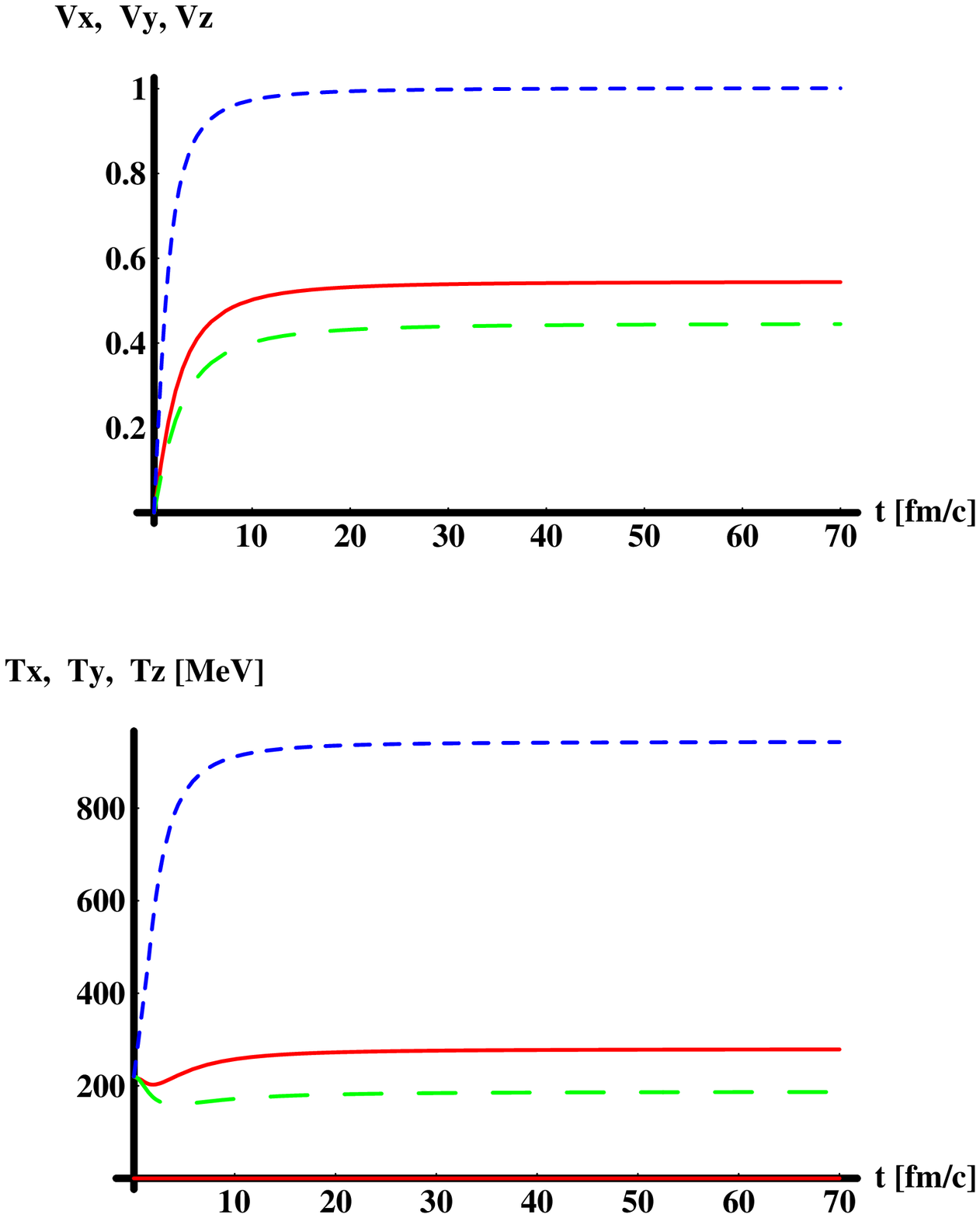}
\caption{
\label{f:exact-timeevol}
Time evolution of an expanding ellipsoid with Gaussian density profile
and homogeneous temperature profile in Buda-Lund type of exact solutions
of non-relativistic hydrodynamics. $(X,Y,Z)$ stand for the principal axis of
this expanding ellipsoid, after an initial acceleration, these scales evolve
linearly with time (top).
The  corresponding HBT radii, $(R_x, R_y, R_z)$
approach a direction independent constant (below top).
The expansion velocities of the principal axis of the exploding
ellipsoid, $(V_x, V_y, V_z)$
tend to direction dependent constants (above bottom panel) and similarly, the
slope parameters of the single particle spectra in the principal directions,
$(T_x, T_y, T_z)$ also tend to direction independent constants (bottom panel).
Based on ref.~\cite{Csorgo:2002kt}.
}
\end{figure}

\begin{figure}
\centerline{\psfig{file=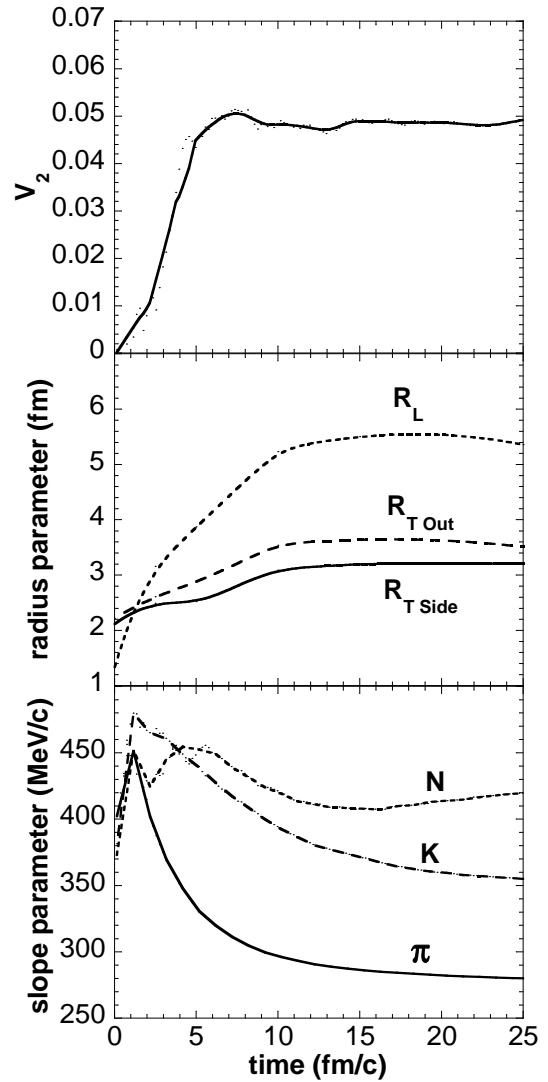,width=7cm}}
\caption{\label{f:timeevol}
Time evolution of $v_2$, HBT, and $m_T$ slope parameters
from HRC for RHIC Au+Au, from ref~\cite{Humanic:2005ye}. The HRC code
shows the same self-quenching properties as the Buda-Lund type of exact
solutions of non-relativistic hydrodynamics.}
\end{figure}

\subsection{Self-consistency criteria - subdivision test}
\label{ss:subdiv}
The self-consistency check of subdivision invariance is based on the
invariance of the Boltzmann equation, the basis of the  Monte Carlo particle-scattering
calculations,  for a simultaneous decrease of the scattering cross sections by some
factor ~$l$, and an increase of the particle density by the same factor of
$l$\cite{Zhang:1998a}. As $l$ becomes sufficiently large,
non-causal artifacts become insignificant. The HRC
rescattering calculations have been tested by comparing pion observables
for no-subdivision, i.e. $l=1$, with subdivision of $l=5$.
Descriptions of how the observables are extracted from the
rescattering calculation are given elsewhere~\cite{Humanic:2002a}.

All calculations were carried out for an impact parameter of 8 fm,
to simulate  semi-central collisions,  that result in significant
elliptic flow. Such HRC results~\cite{Humanic:2005ye} reproduced
reasonably well the RHIC data on the corresponding observables in
Au+Au collisions at $\sqrt{s_{NN}} = 200 $ GeV.   Thus this HRC
code passed not only the test of reproducing the important global
observables, but also the subdivision test, which suggests that
there are no significant programming artifacts, or causality
violating terms in this Monte Carlo simulation code.
Fig.~\ref{f:hrc1} (taken from ref.~\cite{Humanic:2005ye}) shows a
comparisons of the $l=1$ and the $l=5$ cases utilizing the HRC
model to describe the charge averaged pion spectra. Similar plots
were published for the HBT radius parameters and the transverse
momentum and pseudorapidity  $\eta$ dependence of the elliptic
flow. See ref.~\cite{Humanic:2005ye} for further details and for
the comparison plots with RHIC Au+Au data.





\subsection{Self-consistency criteria -
 comparison with  exact hydrodynamical results}
\label{ss:hydro}
Motivated by the success of the Buda-Lund parameterization
~\cite{Csorgo:1995bi,Csanad:2003qa}
of the hadronic final state, a search started to
find parametric, but time dependent, exact solutions of hydrodynamics,
 that lead to Buda-Lund type of nearly Gaussian freeze-out distributions.
Surprisingly large classes of exact, parametric solutions of hydrodynamics
 have been recently found this way, both in the non-relativistic
~\cite{Csorgo:2001ru,Csorgo:2001xm,Csorgo:2002kt}
as well as in the relativistic
~\cite{Biro:1999eh,Csorgo:2003ry,Csorgo:2003rt,Sinyukov:2004am,Csorgo:2006ax,Pratt:2006jj}
kinematic domain.

These generalized, Buda-Lund type parametric, hydrodynamical calculations
have a built-in self-quenching effect, as analyzed in detail in
ref.~\cite{Csorgo:2002kt}.
 For example,
 all the HBT radii stop to evolve in time, they
approach a direction independent constant, and
expansion velocities tend to direction dependent constants, although the system
keeps on expanding (via rescattering process or hydrodynamical evolution).
Elliptic flow also freezes out at the same time when the spectra
(slopes) stop to evolve in time. These general, qualitative properties
of exact parametric hydrodynamic solutions are illustrated in
Fig.~\ref{f:exact-timeevol}. In ref.~\cite{Csorgo:2002kt}
we have shown that this self-quenching is a general
property of a large class of exact analytic hydrodynamic solutions,
which is  independent of the particular initial conditions - a beautiful exact
result. This property of the exact non-relativistic ellipsoidal hydrodynamic
solutions is beautifully reproduced in the Hadronic Resonance Model, as
shown on Fig.~\ref{f:timeevol}.

\subsection{More than hydro - tails of particle production}
\label{ss:more-than-hydro}
As summarized above the HRC resonance cascade model
describes well the observables in heavy ion collisions
at both CERN SPS and at RHIC energies, and it passes
the subdivision test, and yields such a time evolution of
the observables, which is similar to the analytically obtained,
exact hydrodynamical asymptotic behavior of these observables.
So HRC looks to be a reliable model for the production of the
bulk of the particles. In the subsequent part, we investigate
if the HRC model is able to describe also the tails of the
particle production in $\sqrt{s_{NN}} = 200 $
GeV Au+Au collisions.

It turns out that this conventional hadronic cascade model HRC
has a built-in adaptive bin size in the time direction,
hence there is no built-in cutoff time scale in the code. HRC
also contains the cascading of the most abundant
hadrons: $\rho$, $\Delta$, K$^*$, $\omega$, $\eta$, $\eta'$,
$\phi$ and $\Lambda$, but it neglects electrical charge.
Therefore we see that this HRC model implements rescattering in a time
dependent mean free path system, and in the previous section
we have shown that under certain conditions this corresponds to a  random
L\'evy walk and is signaled by power-law tails in the source distribution.
 Let's see whether the HRC simulation results confirm such an
expectation or not.

\section{DETAILED HRC SIMULATION RESULTS}
\label{s:details}

We generated 48 events with b=4.45 fm and 5730 events with 12.5 fm
corresponding to the 0-20\% and 50-90\% centrality classes of
PHENIX events. The generated events have average multiplicities of
3400 and 38 particles, respectively. The mean impact parameter
values suiting the two centralities were determined from a Glauber
calculation as in ref.~\cite{Adler:2003au}.

We made cuts on the data sample similar to the ones in the PHENIX
imaging paper~\cite{PHENIX-imaging}:
\begin{itemize}
\item 0-20\% and  0.2  GeV$/c\;<p_t<\;$0.36 GeV$/c$,
\item 0-20\% and  0.48 GeV$/c\;<p_t<\;$0.6 GeV$/c$,
\item 40-90\% and 0.2  GeV$/c\;<p_t<\;$0.4 GeV$/c$,
\end{itemize}
in addition to the PHENIX geometry cut of $-0.5<y<0.5$.

The resulting plots are shown in
Figs.~\ref{f:hrc_0020_020036_PHENIX}-\ref{f:ptypedep3}.
The complete source function is decomposed to
different components, i.e. the  $S(r)$ distributions
created from pairs with both pions being
\begin{itemize}
\item primordial or decay products of resonances that have a
lifetime less than 20 fm$/c$; this means here direct pions $\rho$, $\Delta$, K$^*$
decay products, referred to as {\it core}, or {\it c} pions;
\item decay products of an $\omega$
\item decay products of resonances
that have a lifetime greater than 25 fm$/c$; in HRC: $\Lambda$,
$\Phi$, $\eta$, $\eta'$, referred to as {\it halo} or {\it h} pions.
\end{itemize}
One of the pions in a pair is from one above category, and the other may be from
another one: so in this study we distinguish $(c,c)$, $(c,\omega)$, $(c,h)$,
$(\omega,\omega)$, $(\omega,h)$ and $(h,h)$ type of possible
combinations.

\subsection{HRC simulations and PHENIX  data}
In this subsection we compare the HRC model calculations with
experimental data from~\cite{PHENIX-imaging}. The kinematic cuts
correspond to Figs. 1a,b and Fig.2 of ref.~\cite{PHENIX-imaging}.

\begin{figure}
  \includegraphics[width=1.0\linewidth]{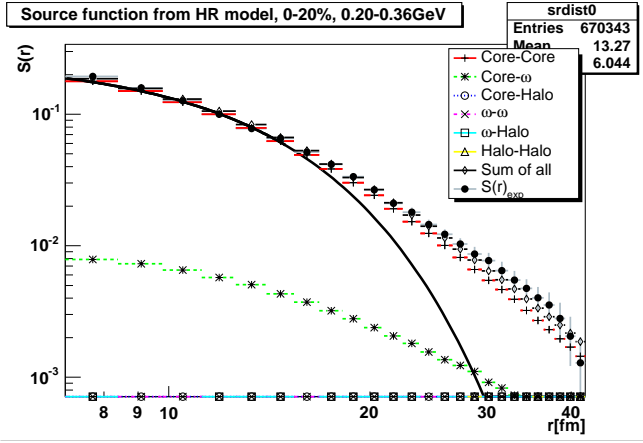}
\caption{Pions for 0-20\% centrality and 0.2 GeV$/c<p_t<$ 0.36 GeV$/c$, with
the various components of the source. On the length scales less
than 50 fm, core-core pairs are the most abundant ones in the
present situation. The next to largest contribution  is from the
$( c,\omega)$
pairs, however, their contribution of
is very small as compared to the core-core pairs
in the experimentally resolvable region, all the other contributions
are negligible.
A comparison of the full HRC simulation result (open diamonds)
with PHENIX data~\cite{PHENIX-imaging}
(filled circles, $S(r)_{exp}$)
indicates that HRC model simulations are in a
good agreement with data in this kinematic range.
Solid black line shows the best
Gaussian fit to $S(r)$ in  the 0 fm $ < r < $ 15 fm region, which
misses the tail region clearly.
\label{f:hrc_0020_020036_PHENIX} }
\end{figure}

The PHENIX imaged $ S(r) $ source distribution coincides with the
HRC simulation result, and both are describable with a power-law
tail, in the kinematic region of 0-20\% centrality and 0.2
GeV$/c<p_t<$ 0.36 GeV$/c$, for pion pairs, as indicated in
Fig.~\ref{f:hrc_0020_020036_PHENIX}. On this log-log plot, the
tail of $S(r)$ is approximately linear, so it can well be called a
\emph{heavy tail}. Clearly a further refinement of the
experimental resolution could reveal more details on the structure
of this tail behavior, but a power-law approximation is not
inconsistent with the currently available data.

In HRC, this power-law tail is due to
rescattering because of the adaptive time-scale, corresponding to
 anomalous diffusion. This results in Levy distributions,
in contrast to Gaussian distributions, which clearly fail to
reproduce the tails of the particle production. The L\'evy index
of stability, $\alpha$, can be determined approximately from fits
to the tails of these $S(r)$ source functions: as for L\'evy
sources $S(r)\propto r^{-(d+\alpha)}$ in $d$ spatial dimensions.
In the HRC simulations, as well as in case of $S(r)$ reconstructed
by PHENIX, $d = 3$. From Fig.~\ref{f:hrc_0020_020036_PHENIX}, the
HRC simulations and PHENIX data both indicate a L\'evy index of
stability of $\alpha \approx 1.15 \pm 0.1$, which is consistent
with the direct L\'evy fits to PHENIX preliminary Coulomb
corrected correlation functions in a similar kinematic region, as
presented in ref.~\cite{Csanad:2005nr}.

This is a great success of the HRC model, as it seems that it has
implemented the key effect, the anomalous diffusion, reasonably well
to the simulation.

\begin{figure}
  \includegraphics[width=1.0\linewidth]{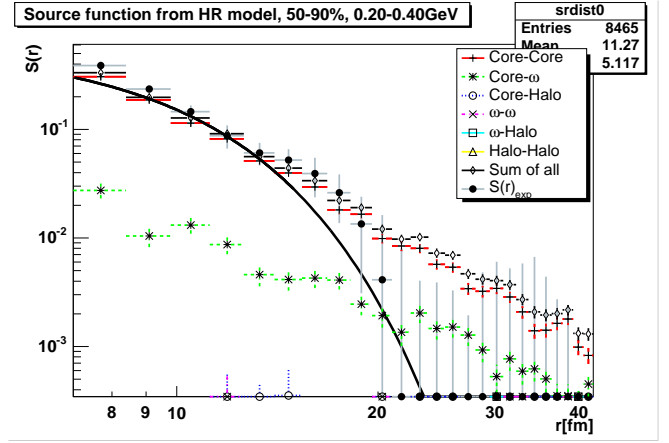}
\caption{Same as the previous figure, but for a
50-90\% centrality and 0.2 GeV$/c<p_t<$ 0.40 GeV$/c$ kinematic selection.
HRC simulations are shown with the various components of the HRC source resolved.
A comparison with PHENIX data~\cite{PHENIX-imaging}
(filled circles, $S(r)_{exp}$) indicates that HRC model simulations are in an
agreement with facts in this kinematic range, too, although
within errors, systematic
deviations between the simulation results and PHENIX data can be observed
in the $r > 20 $ fm region.
\label{f:hrc_5090_020040_PHENIX} }
\end{figure}

Let us investigate in greater detail
if HRC can describe more subtle features of the PHENIX data.
In Fig.~\ref{f:hrc_5090_020040_PHENIX}, we show a comparison
with data in a similarly soft domain but in the more peripheral
centrality class.

Within errors, the HRC simulations again reproduce the PHENIX
measured $S(r)$ source functions, even if the experimental
statistics does not allow for a comparison in the interesting
region of $r>20$ fm. The HRC simulations, which have better
statistics than the experimental data, indicate a power-law tail
with similar exponent as in the case of more central collisions.
From the comparison with the previous figure it follows that the
HRC model describes the centrality dependence of the heavy tail in
this soft $k_t$ region in an acceptable manner. This is yet
another feature of the HRC model which deserves appreciation.

\begin{figure}
  \includegraphics[width=1.0\linewidth]{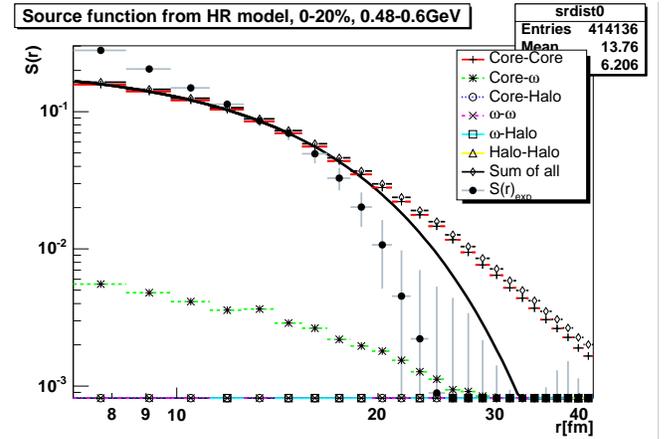}
\caption{Same as the previous figures, but for a
0-20\% centrality and 0.48 GeV$/c<p_t<$ 0.60 GeV$/c$ kinematic selection.
HRC simulations are shown with the various components of the HRC source resolved.
The HRC model simulations are in a
disagreement with PHENIX data in this kinematic range.
\label{f:hrc_0020_048060_PHENIX} }
\end{figure}

Let us now investigate how well can a HRC simulation describe
the transverse momentum dependence of the relative coordinate
distributions of pion pairs. Such a comparison is shown in
Fig.~\ref{f:hrc_0020_048060_PHENIX}.
In the 0-20\% centrality and 0.48 GeV$/c<p_t<$ 0.60 GeV$/c$ kinematic selection,
the HRC model simulations are in a disagreement with PHENIX data.
So the HRC model does not describe the observed transverse momentum
dependence of the PHENIX imaged source distribution in case of nearly central collisions.
Although the achievements of this model are really impressive, in this kinematic
region the model could be improved or fine-tuned. In fact it seems that
rescattering effects are too large as compared to data with increasing transverse
momentum.

In the subsequent parts, we explore the HRC model predictions in
greater detail. We focus our attention to the tails of particle
production, and try to identify within the limitations of this
model calculation what kind of experimental control is available
for changing the exponent (or, on a log-log plot, the slope
parameter) of the tails of particle emission in the HRC
simulations. This is motivated by the recent predictions in
ref.~\cite{2ndQCD} that suggested the existence of a power-law
tail in the coordinate space distribution at the critical end
point of the line of first order phase transitions in QCD. At this
critical point the  phase transition is of second order, and the
Bose-Einstein correlation function has a L\'evy form, with the
L\'evy index of stability coinciding with the correlation exponent
$\eta$ that is one of the critical exponents, and its value is
universal, depending only on the universality class of the second
order phase transition. Hence in a second order QCD phase
transition the L\'evy index of stability or the power-law exponent
of the tails of the particle production becomes independent of the
momentum range, centrality selection, and the particle type.
However, in case of anomalous diffusion of hadrons the
rescattering might well be sensitive to the centrality that drives
multiplicity and particle densities, the momentum range that also
influences how many particles take part in the rescattering
process and also the particle type, as the number of rescatterings
is expected to depend on the particle cross sections.

\subsection{Centrality dependence of the HRC source}
\label{ss:centrality}

\begin{figure}
  \includegraphics[width=1.0\linewidth]{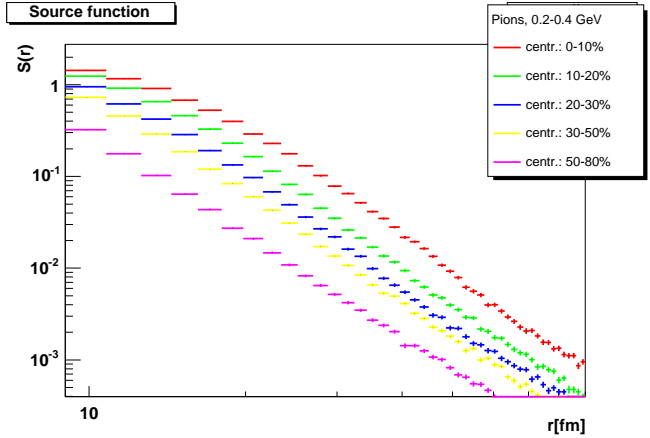}\\
  \caption{Source distribution of HRC simulated pion pairs with
0.2 GeV$/c$ $< p_t <$ 0.4 GeV$/c$ for various centrality classes.\label{f:picent1}
}
\end{figure}

\begin{figure}
  \includegraphics[width=1.0\linewidth]{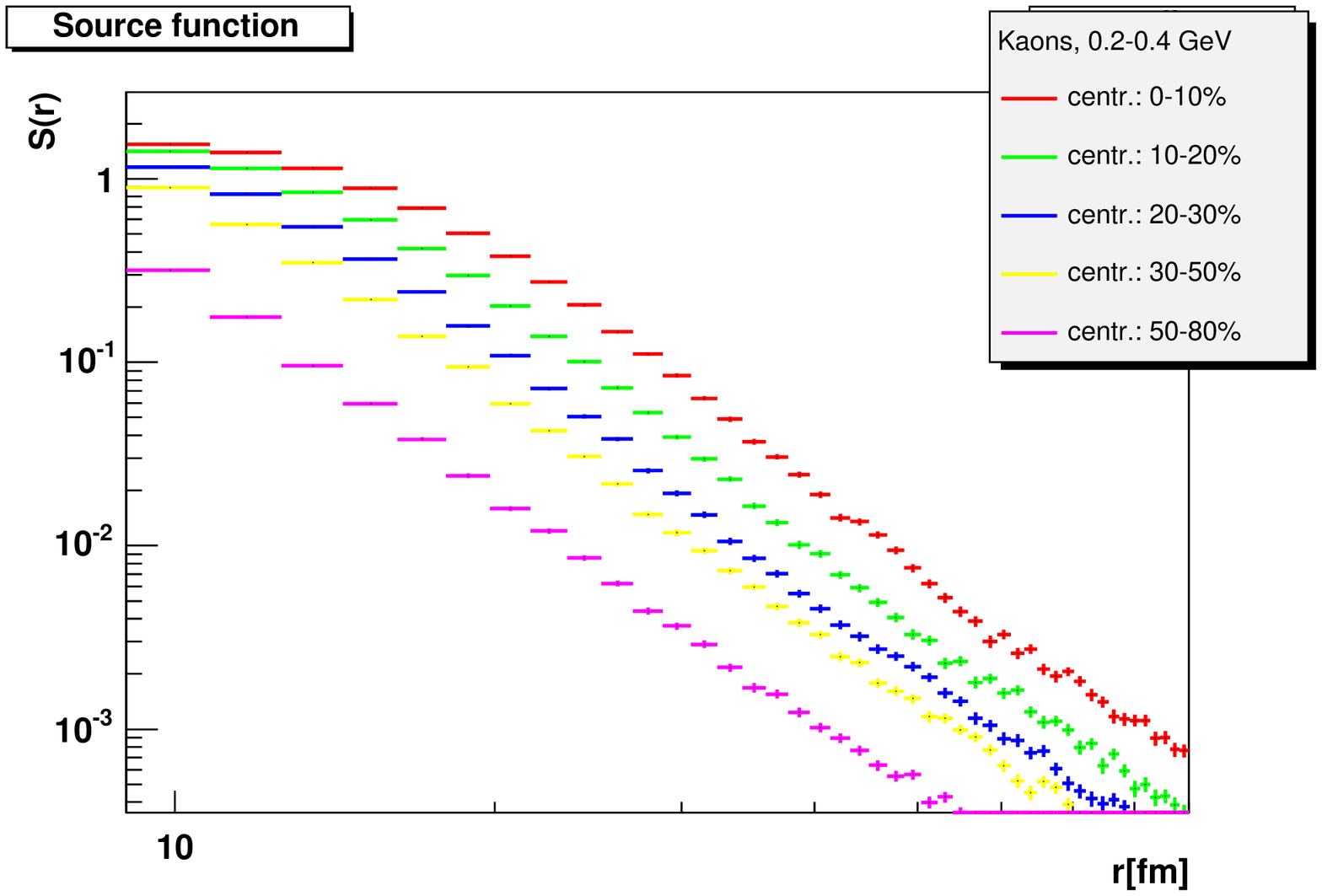}\\
  \caption{Source distribution of HRC simulated pion pairs with
    0.5 GeV$/c$ $< p_t <$ 1.0 GeV$/c$ for various centrality classes.
\label{f:picent2}
}
\end{figure}

\begin{figure}
  \includegraphics[width=1.0\linewidth]{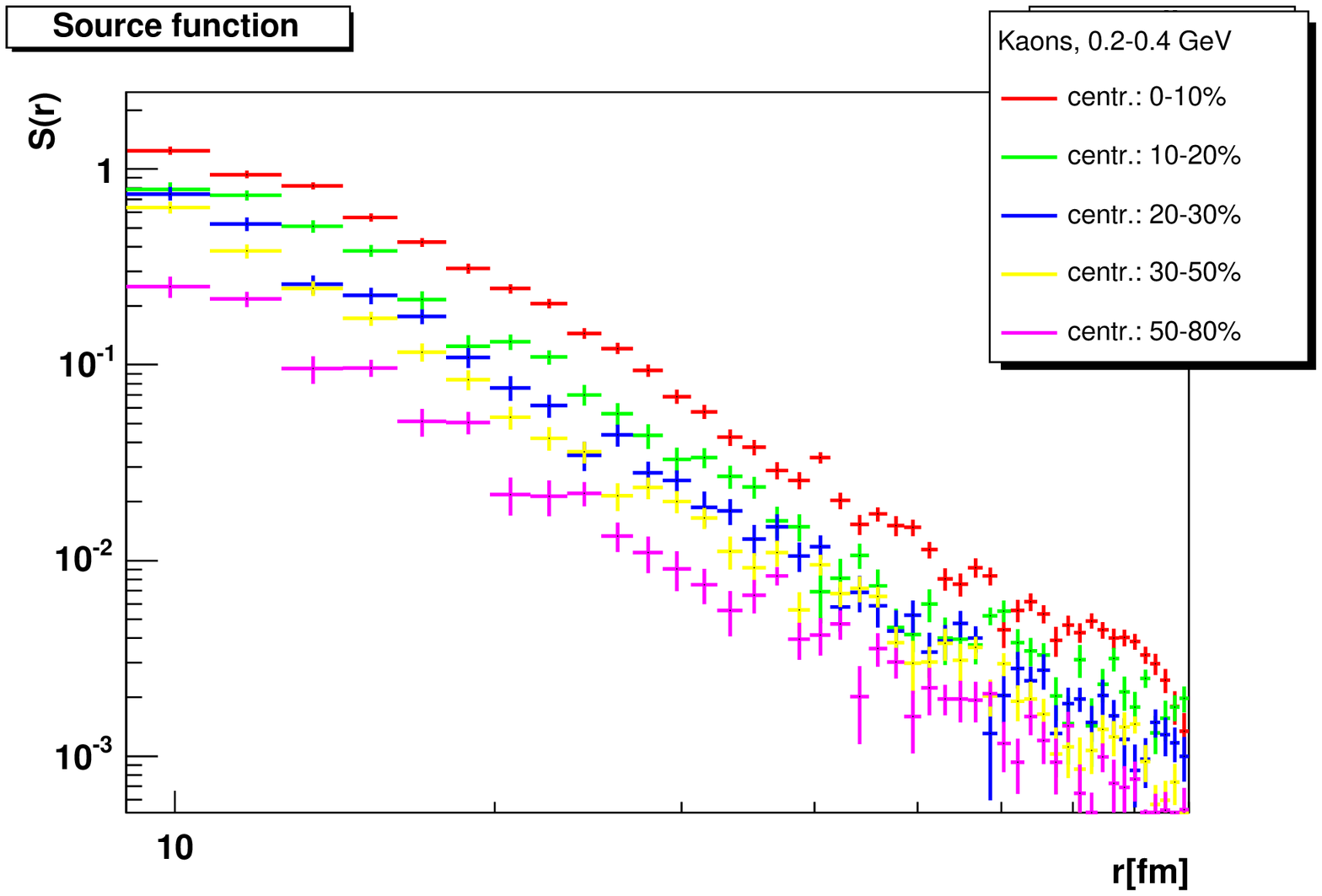}\\
  \caption{Source distribution of kaon pairs with 0.2 GeV$/c$ $< p_t <$ 0.4 GeV$/c$ for various centrality classes.
\label{f:kcent1}
}
\end{figure}
\begin{figure}
  \includegraphics[width=1.0\linewidth]{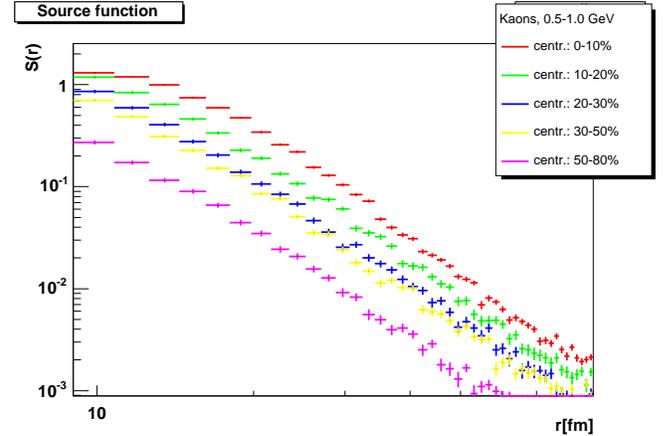}\\
  \caption{Source distribution of kaon pairs with 0.5 GeV$/c$ $< p_t <$ 1.0 GeV$/c$ for various centrality classes.\label{f:kcent2} }
\end{figure}

\begin{figure}
  \includegraphics[width=1.0\linewidth]{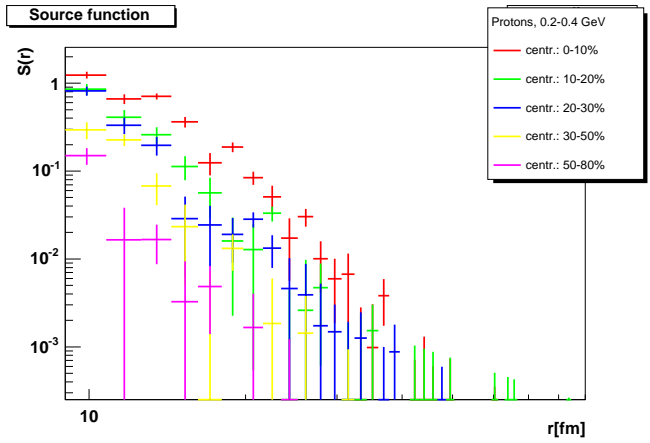}\\
  \caption{Source distribution of proton pairs with 0.2 GeV$/c$ $< p_t <$ 0.4 GeV$/c$ for various centrality classes.\label{f:prcent1} }
\end{figure}

\begin{figure}
  \includegraphics[width=1.0\linewidth]{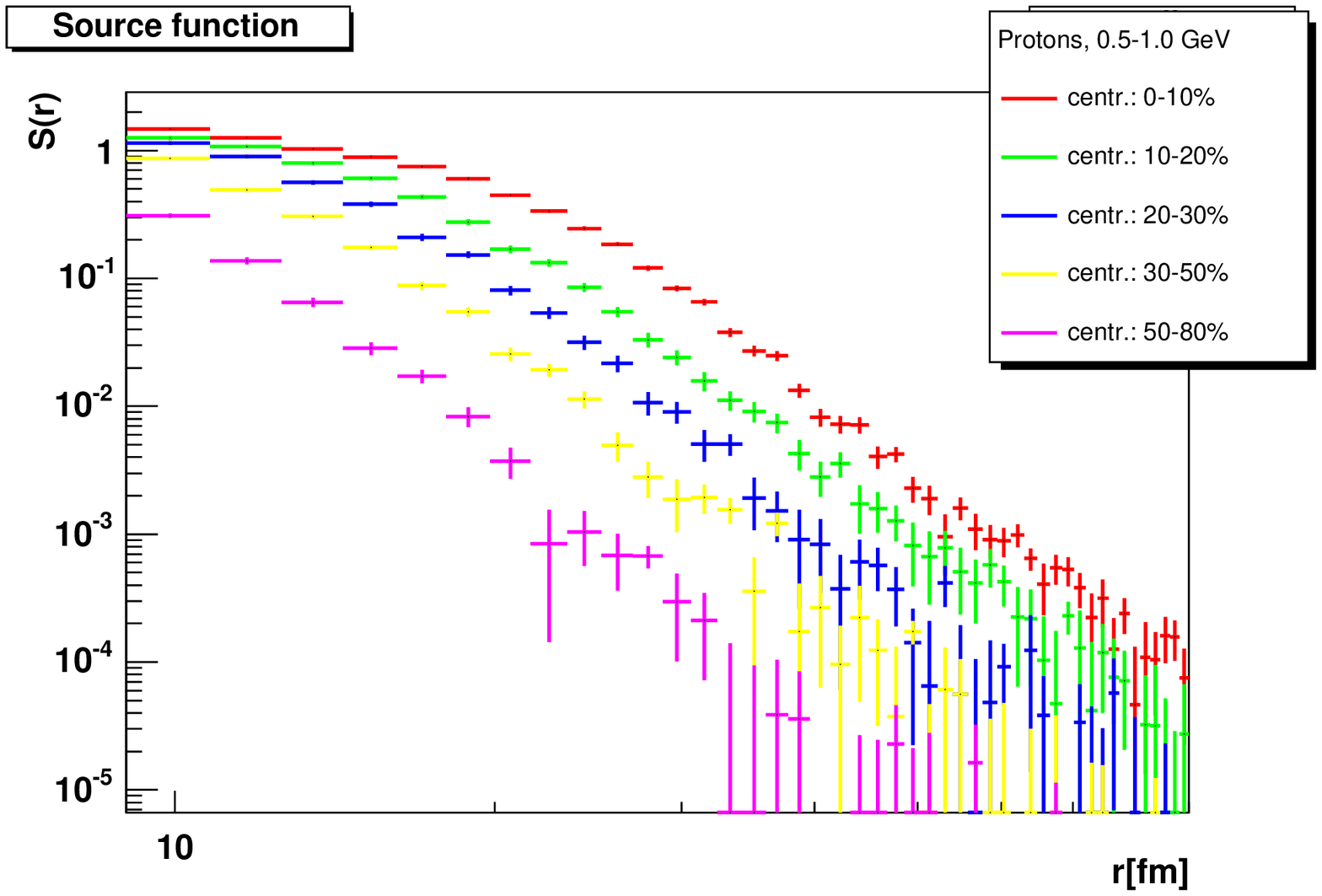}\\
  \caption{Source distribution of proton pairs with 0.5 GeV$/c$ $< p_t <$ 1.0 GeV$/c$ for various centrality classes.\label{f:prcent2} }
\end{figure}

The centrality dependence of the HRC simulated source
distributions of pions with various pair transverse momenta is
shown in Figs.~\ref{f:picent1}-\ref{f:prcent2}. The centrality
dependence of the tails of particle emissions is found to be
surprisingly small, as on the log-log plots the tails
corresponding to various centrality selections in
Figs.~\ref{f:picent1}-\ref{f:prcent2} are found to be parallel.
The centrality selection, however, sensitively influences the
region of bulk particle production, corresponding to the change of
the scale parameters or the effective source sizes in the small
$r$ region. The slope of the tails in this simulation, however, is
independent of centrality for pions at low or higher transverse
momentum, Figs.~\ref{f:picent1}-\ref{f:picent2}, as well as for
low or higher momentum kaons, Figs.~\ref{f:kcent1}-\ref{f:kcent2},
and for protons, Figs.~\ref{f:prcent1}-\ref{f:prcent2}. Although
in case of the proton sources statistical limitations prevent us
from firm conclusion, it seems that the power-law exponent of the
tails of particle emission is rather insensitive to the centrality
selection, regardless of particle type and pair momentum
selection,
 so centrality is not  a sensitive control
tool to separate heavy tails that arise due to anomalous diffusion
from heavy tails that appear due to the vicinity of
a second order QCD phase transition.


\subsection{Transverse momentum dependence}
\label{ss:kt}

\begin{figure}
  \includegraphics[width=1.0\linewidth]{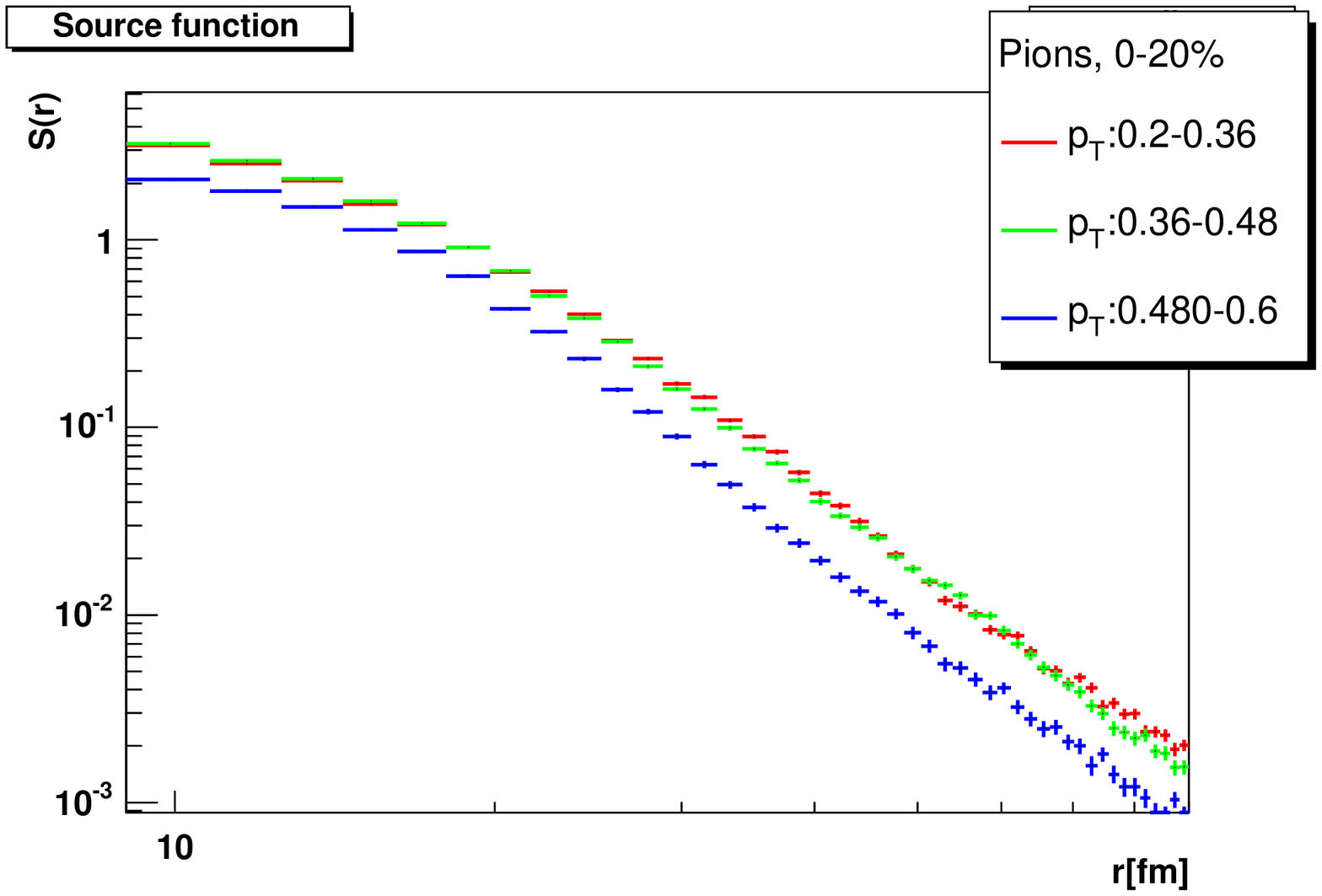}\\
  \caption{Source distribution of pion pairs with 0-20\% centrality,
    for various $p_t$ ranges}\label{f:pipt1}
\end{figure}
\begin{figure}
  \includegraphics[width=1.0\linewidth]{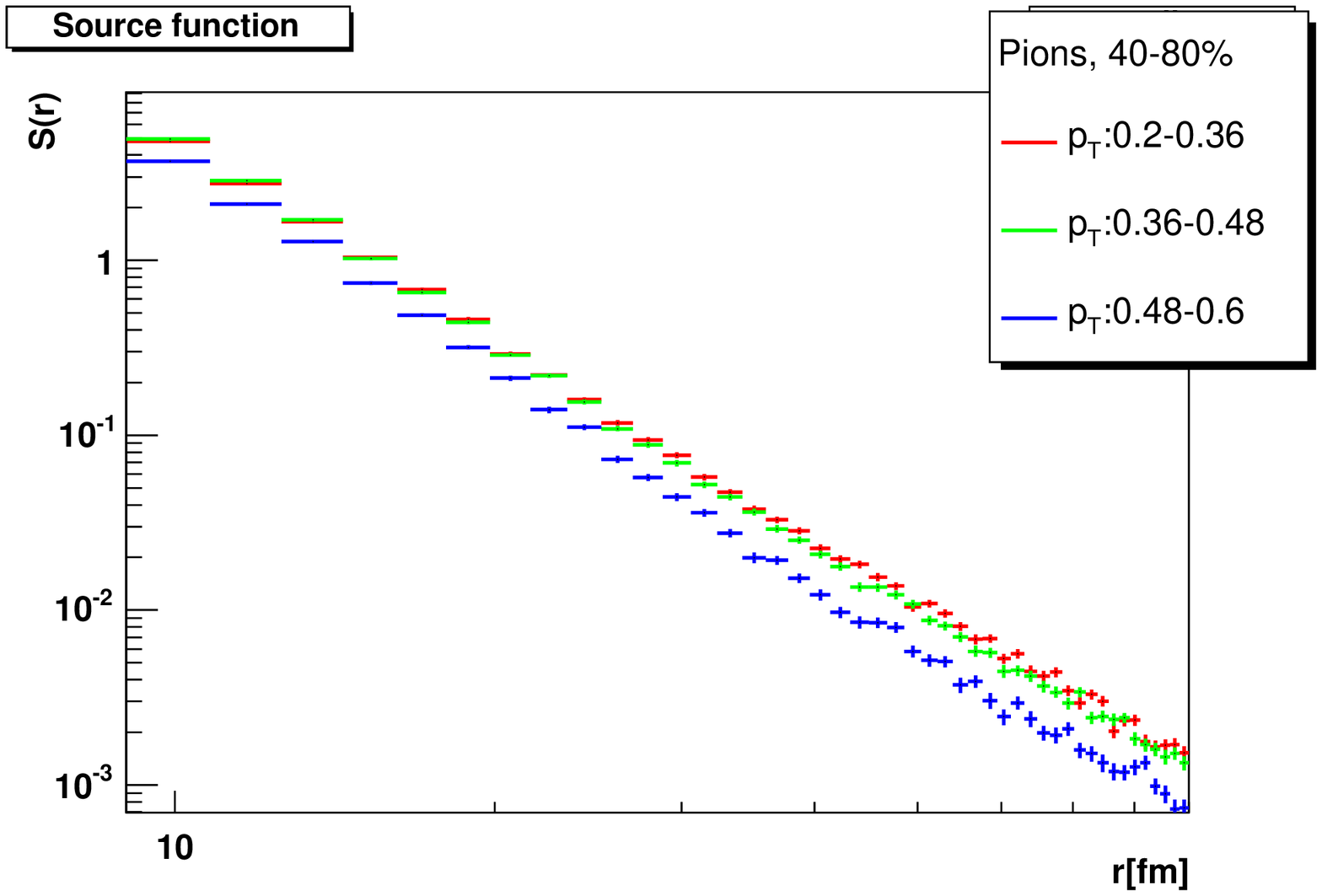}\\
  \caption{Source distribution of pion pairs with 40-80\% centrality,
    for various $p_t$ ranges}\label{f:pipt2}
\end{figure}

\begin{figure}
  \includegraphics[width=1.0\linewidth]{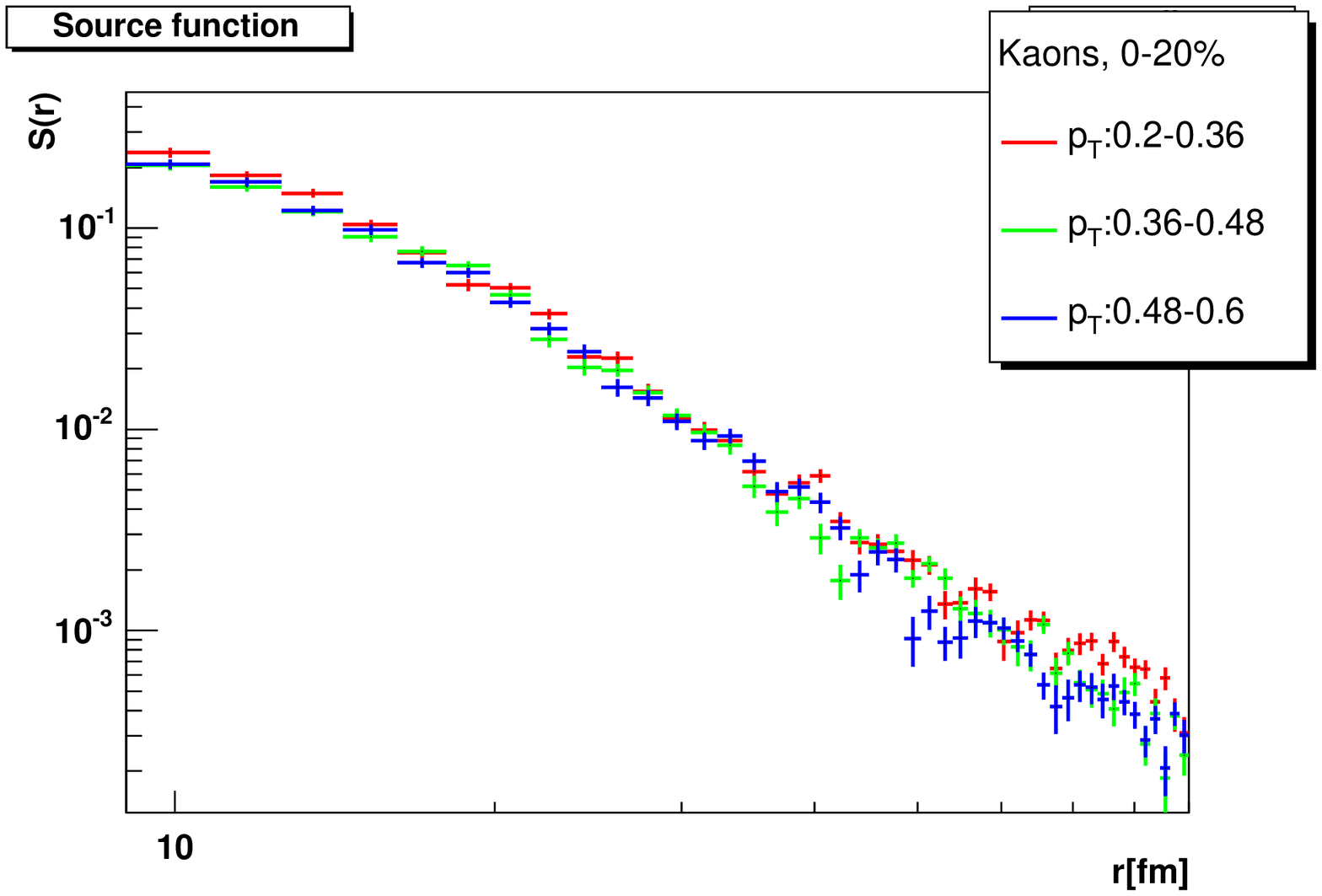}\\
  \caption{Source distribution of kaon pairs with 0-20\% centrality,
    for various $p_t$ ranges}\label{f:kpt1}
\end{figure}
\begin{figure}
  \includegraphics[width=1.0\linewidth]{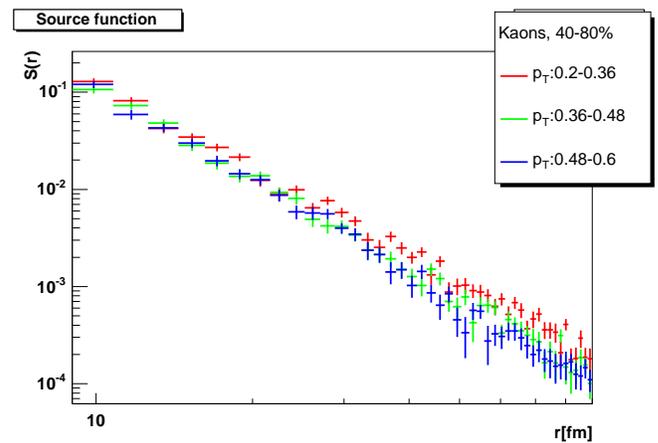}\\
  \caption{Source distribution of kaon pairs with 40-80\% centrality,
    for various $p_t$ ranges}\label{f:kpt2}
\end{figure}

\begin{figure}
  \includegraphics[width=1.0\linewidth]{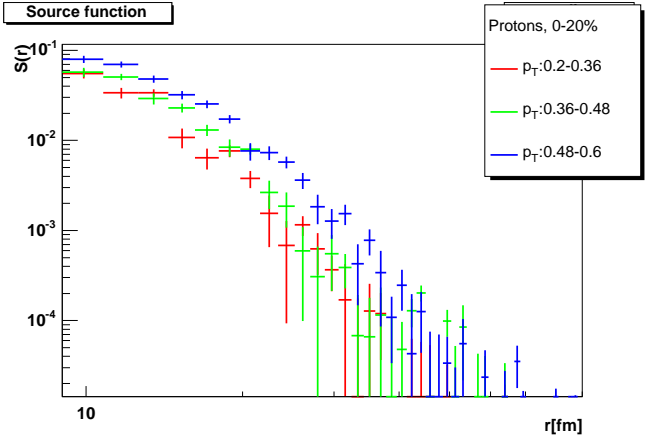}\\
  \caption{Source distribution of proton pairs with 0-20\% centrality,
    for various $p_t$ ranges}\label{f:prpt1}
\end{figure}

\begin{figure}
  \includegraphics[width=1.0\linewidth]{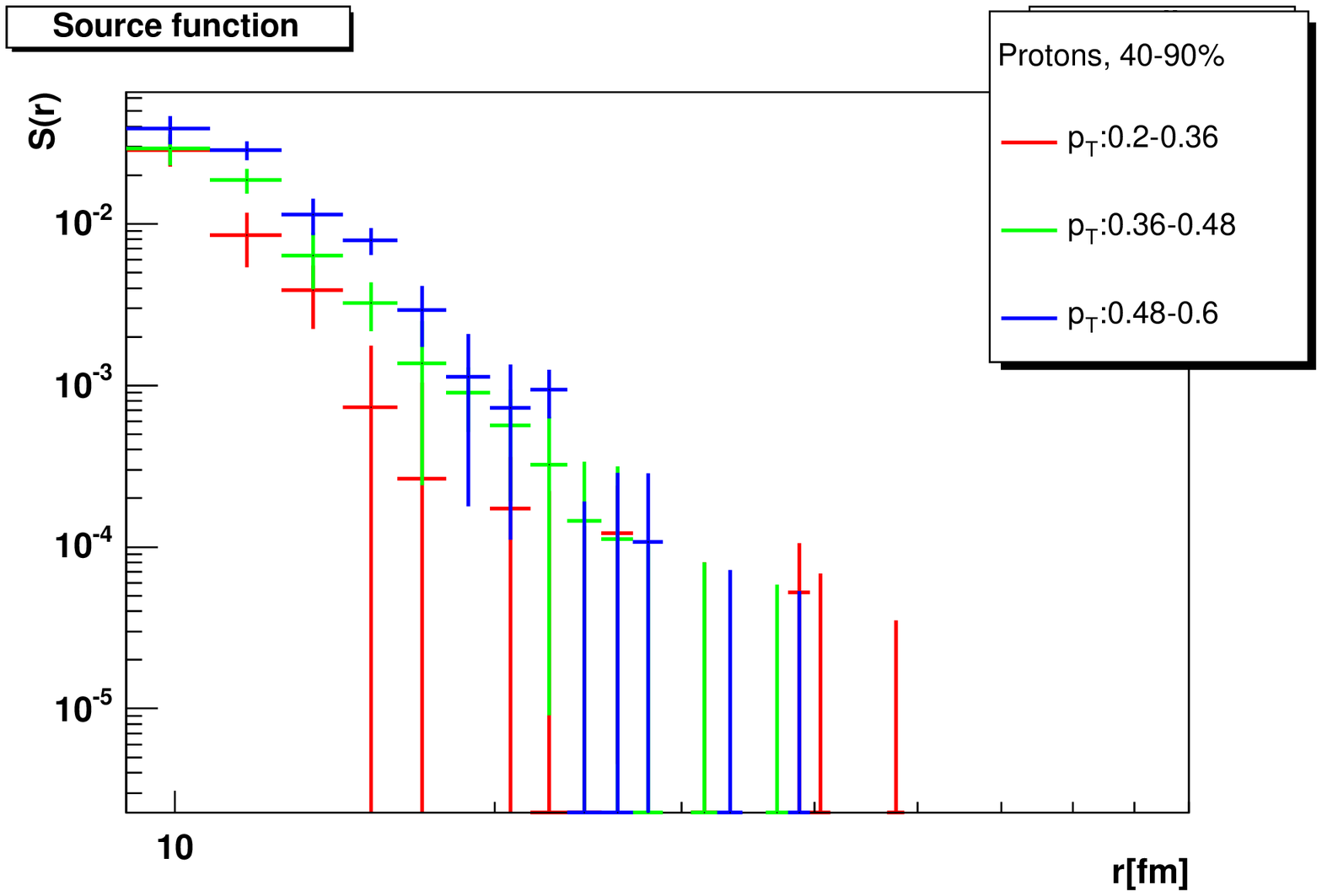}\\
  \caption{Source distribution of proton pairs with 40-80\% centrality,
    for various $p_t$ ranges}\label{f:prpt2}
\end{figure}

\begin{figure}
  \includegraphics[width=1.0\linewidth]{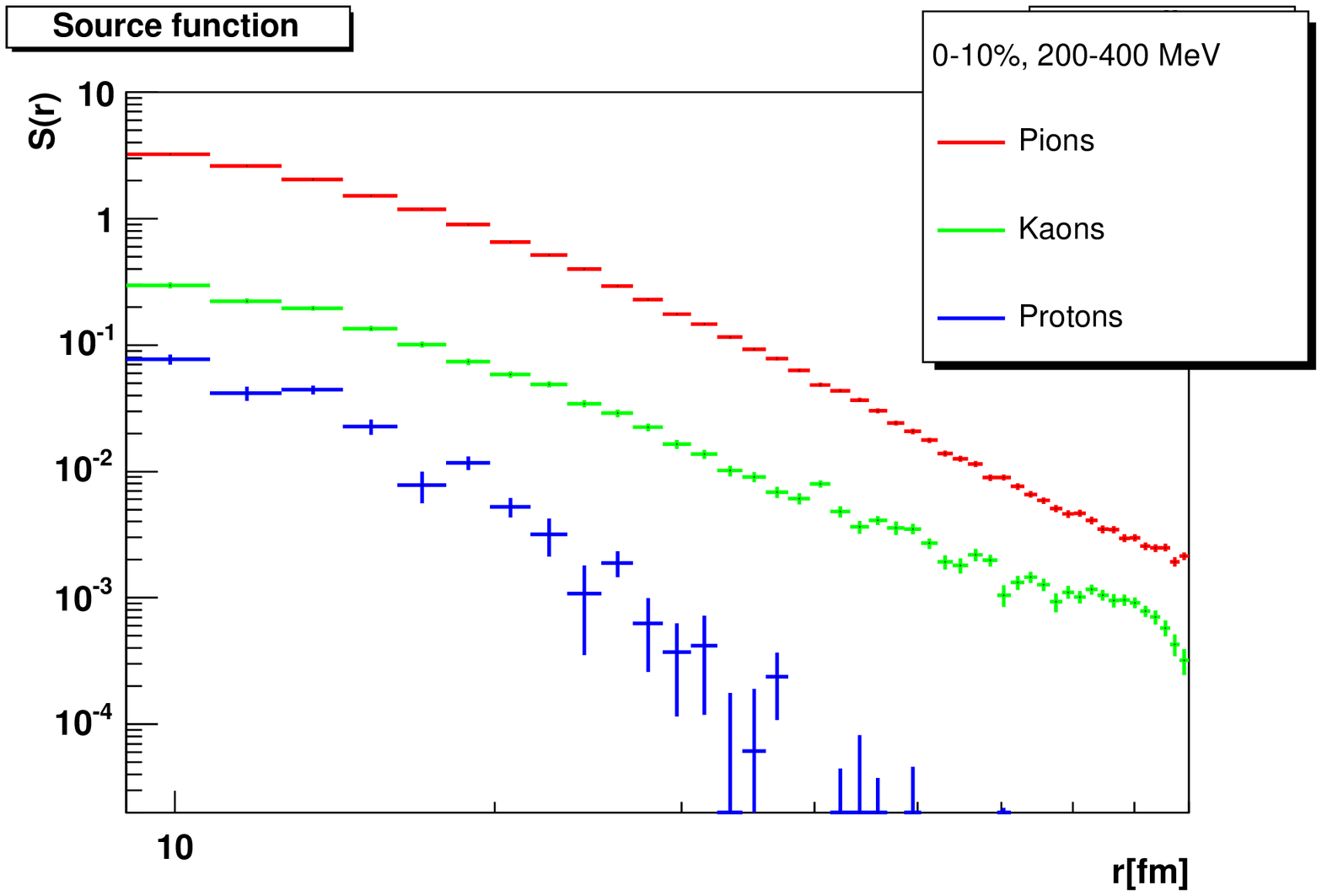}
  \caption{Source distribution of pion, kaon and proton pairs with
  0.2 GeV$/c$ $< p_t <$ 0.4 GeV$/c$ and 0-10\% centrality.}\label{f:ptypedep1}
\end{figure}

\begin{figure}
  \includegraphics[width=1.0\linewidth]{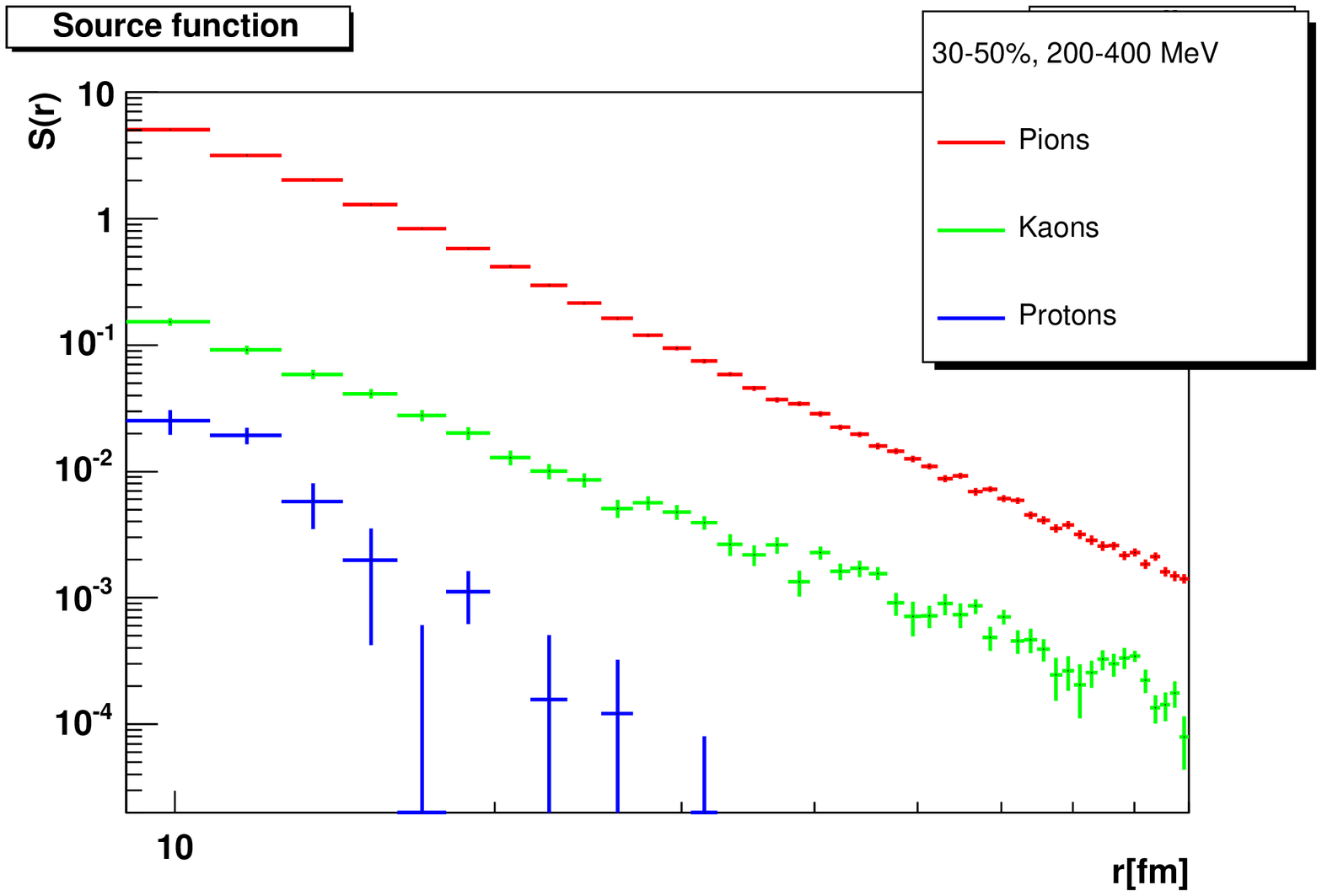}
  \caption{Source distribution of pion, kaon and proton pairs with
  0.2 GeV$/c$ $< p_t <$ 0.4 GeV$/c$ and 30-50\% centrality.}\label{f:ptypedep2}
\end{figure}

\begin{figure}
  \includegraphics[width=1.0\linewidth]{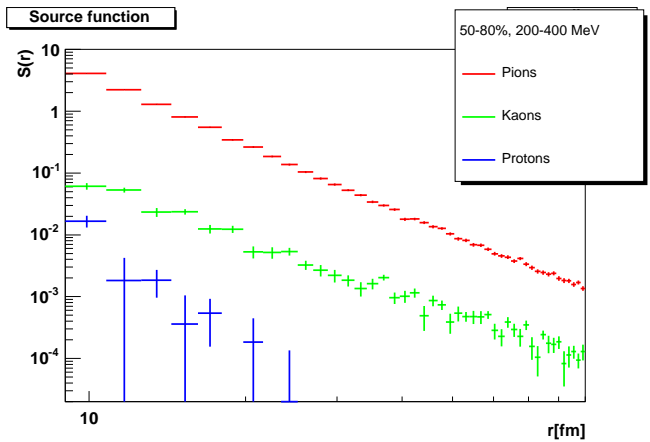}
  \caption{Source distribution of pion, kaon and proton pairs with
  0.2 GeV$/c$ $< p_t <$ 0.4 GeV$/c$ and 50-80\% centrality}\label{f:ptypedep3}
\end{figure}

Transverse momentum dependence of simulated source distributions
in various centrality classes is shown in
Figs.~\ref{f:pipt1}-\ref{f:prpt2}.
For nearly central collisions and pions, the increase of the transverse
momentum of the pair reduces the size of the bulk production region,
as was well known from earlier Bose-Einstein correlation measurements and
theoretical explanations, see e.g.
refs.~\cite{Csorgo:1995bi,Csanad:2003qa}.
Being aware of the strong transverse momentum dependences of the scale parameter
$R$, which is well shown by the HRC simulation, it is rather surprising,
that the shape parameter of the tail, the L\'evy index of stability $\alpha$
is remarkably insensitive to the selected transverse momentum regions,
which follows from the fact that on the log-log plot the simulated $S(r)$
tails are parallel to one another, see Fig.~\ref{f:pipt1}.
The same effect is shown in Fig.~\ref{f:pipt2}, in case of peripheral collisions.
For kaons, the evolution of the source function is less apparent in the
same transverse momentum regions, this is due to the fact that the
effective radius parameters, the scales in analytic calculations
like of refs.~\cite{Csorgo:1995bi,Csanad:2003qa}, are found to be
predominantly depending on the transverse mass, $m_t = \sqrt{m^2 + p_t^2}$
of the particles.
The same variation in the transverse momentum  $p_t$ leads to a larger
variation in    $m_t$ for pions, as compared to that of kaons, due to
the larger value of the kaon mass: $m_{\pi} \approx  140$ MeV,
while $m_K \approx 494 $ MeV. This explains why
Figs.~\ref{f:kpt1} and ~\ref{f:kpt2} show remarkably small variations
of the kaon emitting source with increasing values of the transverse momenta
of kaons. In case of protons, one would expect even smaller variations
with increasing momentum, however, in this case the Monte Carlo simulations
indicate an increase of the effective source size with increasing momentum.
Even in case of protons the dependence of the power-law exponent on the transverse
momentum is negligible.  Thus the transverse momentum dependence of the power-law
exponent $\alpha$ is
negligible in each considered case.

\subsection{Particle type dependence}
\label{ss:pid}

Particle type dependence of simulated source distributions in
various centrality classes and with a pair transverse momentum of
0.2 GeV$/c$ $< p_t <$ 0.4 GeV$/c$ is shown in
Figs.~\ref{f:pipt1}-\ref{f:prpt2}. These source distributions all
have an approximate power-law tail, as expected in case of an anomalous
diffusion effect. However, the power-law exponents have rather
different values for different particle types. This is well understandable
as the total inelastic  cross sections of these particles are rather different.
Protons have the largest cross sections, pions the second largest,
and kaons the smallest. Hence the protons have the shortest mean free path,
the pions the second shortest, and the kaons have the largest mean free path
at any given densities. This implies that the heaviest tail develops for kaons,
the second heaviest tail for pions, and the proton source distribution is
closest to the Gaussian distribution. However, it is rather difficult to make
such a simple picture given that the HRC simulation includes a parameterized
form of the strongly momentum dependent cross sections, when available,
it uses Particle Data Group tables~\cite{PDG92}, although not the most recent
version of these tables~\cite{PDG06}.
Whenever PDG data were not available,
the HRC model utilized theoretical results on the total cross sections from Prakash, Prakash,
Venugopalan and Welke~\cite{Prakash:1993bt}.

\section{CONCLUSIONS AND SUMMARY}
\label{s:concl}

A power-law tail of the relative coordinate distribution $S(r)$ appears in HRC model simulations for Au+Au collisions
at RHIC energies. The simulation reproduces an important
feature of recent PHENIX experimental data on heavy tails, i.e. it
yields a power-law tail with an exponent, or L\'evy index of stability
of $\alpha \approx 1.3 \pm 0.1$. Detailed simulation
results indicate that this exponent (as well as the tails of the
$S(r)$ relative coordinate distribution) depends strongly on the
particle cross sections or particle type, while for a given particle type
it is remarkably insensitive to centrality and momentum selections, which
implies insensitivity to the actual number of collisions.
The exponent value is rather different from $\alpha \approx 0.5$
predicted for the second order QCD phase transition in ref.~\cite{2ndQCD}.
Also, at a second order QCD phase transition, the exponents
are determined from universality class arguments, so in case
of a second order phase transition, no particle type dependence
is expected. Hence measuring the tails of particle production for protons
and kaons seems to be a promising method to distinguish between
a second order QCD phase transition and anomalous diffusion.

The L\'evy distribution is called stable because the convolution
of two such distribution is also a L\'evy distribution. In addition,
the generalized central limit theorem says that the convolution of
many (in the limiting case infinitely many) elementary processes
with the same (but arbitrary) probability distribution
is a L\'evy stable distribution. So it is natural, although
still surprising that L\'evy stable, power-law tailed source
distributions appear in the HRC model simulations. (The
rescattering can be considered as such elementary process.)

The extraction of the precise values of the power-law exponents,
or the L\'evy index of stability, with reliable experimental
statistical and systematic errors is an important future task
that allows experimental characterization of anomalous diffusion
or second order QCD phase transitions with a simple number. Measuring the
transverse momentum, centrality, and particle type dependence of this exponent
tells a lot about the production mechanism  of particles in high energy
heavy ion collisions.  In particular, measuring
the L\'evy index of stability for pions, kaons, and protons
can serve as a promising experimental control possibility
on the origin of the heavy tailed distribution.

\bigskip
\noindent\textbf{Acknowledgments:}
We are grateful to professor Tom Humanic
for forwarding his HRC simulation code to us.
T. Cs. would like to express his deep gratitude to professors S.S. Padula,
Y. Hama and M. Hussein for the kind invitation and hospitality
and for their  organizing a series of  outstanding conferences in Brazil.


\begin{thebibliography}{99}
\bibitem{Lednicky:2005af}
  R.~Lednicky,
  Nucl.\ Phys.\ A {\bf 774} (2006) 189
  [arXiv:nucl-th/0510020].

\bibitem{Padula:2004ba}
  S.~S.~Padula,
  Braz.\ J.\ Phys.\  {\bf 35}, 70 (2005)
  [arXiv:nucl-th/0412103].

\bibitem{Csorgo:2005gd}
  T.~Cs\"org\H{o},
  J.\ Phys.\ Conf.\ Ser.\  {\bf 50}, 259 (2006)
  [arXiv:nucl-th/0505019].

\bibitem{Lisa:2005dd}
  M.~A.~Lisa, S.~Pratt, R.~Soltz and U.~Wiedemann,
  Ann.\ Rev.\ Nucl.\ Part.\ Sci.\  {\bf 55}, 357 (2005)
  [arXiv:nucl-ex/0505014].

\bibitem{Lisa:2005js}
  M.~Lisa,
  AIP Conf.\ Proc.\  {\bf 828}, 226 (2006)
  [arXiv:nucl-ex/0512008].


\bibitem{Hwa:2007rp}
  R.~C.~Hwa,
  arXiv:nucl-th/0701053.

\bibitem{PHENIX-imaging}
  S.~S.~Adler {\it et al.}  [PHENIX Collaboration],
  arXiv:nucl-ex/0605032.

\bibitem{Bialas:1992ca}
  A.~Bia{\l}as,
  Acta Phys.\ Polon.\ B {\bf 23}, 561 (1992).

\bibitem{MK-rep}
    R. Metzler, J. Klafter, Physics Reports 339 (2000) 1-77

\bibitem{Humanic:2005ye}
  T.~J.~Humanic,
  Int.\ J.\ Mod.\ Phys.\ E {\bf 15}, 197 (2006)
  [arXiv:nucl-th/0510049].

\bibitem{zol2}
        V. V. Uchaikin and V. M. Zolotarev,
        {\it ``Chance and Stability, Stable Distributions and
        Their Applications"}, VSP Science,1999, ISBN: 90-6764-301-7, 596 pp.

\bibitem{zol1}
        V. M. Zolotarev, {\it ``One-dimensional Stable Distributions"},
        {\it  Am. Math. Soc. Transl. of Math. Monographs}, vol. {\bf 65},
        Providence, R.I. (Transl. of the original 1983 Russian)


\bibitem{nolan-chap1}
        J. P. Nolan, {\it Stable Distributions: Models for Heavy Tailed Data}
        \hfill\\
        {\tt\scriptsize http://academic2.american.edu/$\tilde{\,\,\,}$jpnolan/stable/CHAP1.PDF}

\bibitem{nolan-book}
    Nolan, J. P. Stable Distributions: Models for Heavy Tailed Data.
     Boston, MA: Birkhäuser, 2005

\bibitem{csorgo-hegyi-zajc}
  T.~Cs\"org\H{o}, S.~Hegyi and W.~A.~Zajc,
  Eur.\ Phys.\ J.\ C {\bf 36}, 67 (2004)
  [arXiv:nucl-th/0310042].

\bibitem{STAR-edge}
  J.~Adams {\it et al.}  [STAR Collaboration],
  Phys.\ Rev.\ C {\bf 71}, 044906 (2005)
  [arXiv:nucl-ex/0411036].



\bibitem{imaging}
  D.~A.~Brown and P.~Danielewicz,
  Phys.\ Lett.\ B {\bf 398}, 252 (1997)
  [arXiv:nucl-th/9701010].




\bibitem{Humanic:2003gs}
  T.~J.~Humanic,
  arXiv:nucl-th/0301055.

\bibitem{Zhang:1999bd}
  B.~Zhang, C.~M.~Ko, B.~A.~Li and Z.~w.~Lin,
  Phys.\ Rev.\ C {\bf 61}, 067901 (2000)
  [arXiv:nucl-th/9907017].


\bibitem{Humanic:2002a}
T.~J.~Humanic,
arXiv:nucl-th/0205053.

\bibitem{Ko:2002iz}
  C.~M.~Ko, Z.~W.~Lin and S.~Pal,
  Heavy Ion Phys.\  {\bf 17}, 219 (2003)
  [arXiv:nucl-th/0205056].



\bibitem{Zhang:1998a}
B.~Zhang, M.~Gyulassy and Y.~Pang,
Phys.\ Rev.\ C {\bf 58}, 1175 (1998)

\bibitem{Csorgo:1995bi}
  T.~Cs\"org\H{o} and B.~L\"orstad,
  Phys.\ Rev.\ C {\bf 54}, 1390 (1996)
  [arXiv:hep-ph/9509213].

\bibitem{Csanad:2003qa}
  M.~Csan\'ad, T.~Cs\"org\H{o} and B.~L\"orstad,
  Nucl.\ Phys.\ A {\bf 742}, 80 (2004)
  [arXiv:nucl-th/0310040].

\bibitem{Csorgo:2001ru}
  T.~Cs\"org\H{o},
  Acta Phys.\ Polon.\ B {\bf 37}, 483 (2006)
  [arXiv:hep-ph/0111139].

\bibitem{Csorgo:2001xm}
  T.~Cs\"org\H{o}, S.~V.~Akkelin, Y.~Hama, B.~Luk\'acs and Yu.~M.~Sinyukov,
  Phys.\ Rev.\ C {\bf 67}, 034904 (2003)
  [arXiv:hep-ph/0108067].

\bibitem{Csorgo:2002kt}
  T.~Cs\"org\H{o} and J.~Zim\'anyi,
  Heavy Ion Phys.\  {\bf 17}, 281 (2003)
  [arXiv:nucl-th/0206051].

\bibitem{Biro:1999eh}
  T.~S.~Bir\'o,
  Phys.\ Lett.\ B {\bf 474}, 21 (2000)
  [arXiv:nucl-th/9911004].

\bibitem{Csorgo:2003ry}
  T.~Cs\"org\H{o}, L.~P.~Csernai, Y.~Hama and T.~Kodama,
  Heavy Ion Phys.\ A {\bf 21}, 73 (2004)
  [arXiv:nucl-th/0306004].

\bibitem{Csorgo:2003rt}
  T.~Cs\"org\H{o}, F.~Grassi, Y.~Hama and T.~Kodama,
  Phys.\ Lett.\ B {\bf 565}, 107 (2003)
  [arXiv:nucl-th/0305059].

\bibitem{Sinyukov:2004am}
  Yu.~M.~Sinyukov and I.~A.~Karpenko,
  Acta Phys.Hung. {\bf A25}, 141-147 (2006)
    [arXiv:nucl-th/0506002].

\bibitem{Csorgo:2006ax}
  T.~Cs\"org\H{o}, M.~I.~Nagy and M.~Csan\'ad,
  arXiv:nucl-th/0605070.

\bibitem{Pratt:2006jj}
  S.~Pratt,
  arXiv:nucl-th/0612010.

\bibitem{Adler:2003au}
  S.~S.~Adler {\it et al.}  [PHENIX Collaboration],
        Phys.\ Rev.\  C {\bf 69}, 034910 (2004)
      [arXiv:nucl-ex/0308006].

\bibitem{Csanad:2005nr}
  M.~Csanad  [PHENIX Collaboration],
  Nucl.\ Phys.\ A {\bf 774}, 611 (2006)
  [arXiv:nucl-ex/0509042].


\bibitem{2ndQCD}
  T.~Cs\"org\H{o}, S.~Hegyi, T.~Nov\'ak and W.~A.~Zajc,
  AIP Conf.\ Proc.\  {\bf 828}, 525 (2006)
  [arXiv:nucl-th/0512060]; \\
  T.~Cs\"org\H{o}, S.~Hegyi, T.~Nov\'ak and W.~A.~Zajc,
  Acta Phys.\ Polon.\ B {\bf 36}, 329 (2005)
  [arXiv:hep-ph/0412243].

\bibitem{PDG92}
    K. Hikasa {\it et al.}, Particle Data Group,
    Phys. Rev. D {\bf 45} (1992) 83

\bibitem{PDG06}
    W-M Yao {\it et al.}, Particle Data Group,
     J. Phys. G: Nucl. Part. Phys. {\bf 33} (2006) 1-1232

\bibitem{Prakash:1993bt}
  M.~Prakash, M.~Prakash, R.~Venugopalan and G.~Welke,
  Phys.\ Rept.\  {\bf 227}, 321 (1993).

\end{thebibliography}
\end{document}